\documentclass{aa}

\usepackage{hyperref}
\usepackage{txfonts}
\usepackage{graphicx}

\usepackage{natbib}
\bibpunct{(}{)}{;}{a}{}{,} 

\begin{document}

\title{Towards a consistent framework of comparing galaxy mergers in observations and simulations}

\author{L.~Wang\inst{1,2}\and W.~J.~Pearson\inst{1,2,3}\and V. Rodriguez-Gomez\inst{4}}

\institute{SRON Netherlands Institute for Space Research, Landleven 12, 9747 AD, Groningen, The Netherlands \email{l.wang@sron.nl} 
\and Kapteyn Astronomical Institute, University of Groningen, Postbus 800, 9700 AV Groningen, the Netherlands
\and National Centre for Nuclear Research, ul. Pasteura 7, 02-093 Warszawa, Poland
\and Instituto de Radioastronom\'ia y Astrof\'isica, Universidad Nacional Aut\'onoma de M\'exico, Apdo. Postal 72-3, 58089 Morelia, Mexico
}
\date{Received / Accepted}

\abstract
   {}
   {We aim to perform consistent comparisons between observations and simulations on the mass dependence of the galaxy major merger fraction at low redshift over an unprecedentedly wide range of stellar masses ($\sim10^{9}$ to $10^{12} M_{\odot}$).}
   {We first carry out forward modelling of ideal synthetic images of major mergers and non-mergers selected from the Next Generation Illustris Simulations (IllustrisTNG) to include major observational effects. We then train deep convolutional neural networks (CNNs) using realistic mock observations of galaxy samples from the simulations. Subsequently, we apply the trained CNNs to real the Kilo-Degree Survey (KiDS) images of galaxies selected from the Galaxy And Mass Assembly (GAMA) survey. Based on the major merger samples, which are detected in a consistent manner in the observations and simulations, we determine the dependence of major merger fraction on stellar mass at $z\sim0.15$ and make comparisons between the two.}
   { The detected major merger fraction in the GAMA/KiDS observations has a fairly mild decreasing trend with increasing stellar mass over the mass range $10^{9}M_{\odot}<M_*<10^{11.5}M_{\odot}$. There is good agreement in  the mass dependence of the major merger fraction  in the GAMA/KiDS observations and the IllustrisTNG simulations over $10^{9.5}M_{\odot}<M_*<10^{10.5}M_{\odot}$. However,  the observations and the simulations show some differences at $M_*>10^{10.5}M_{\odot}$, possibly due to  the supermassive blackhole feedback in its low-accretion state in the simulations which causes a sharp transition in the quenched fractions at this mass scale. The discrepancy could also be due to the relatively small volume of the simulations and/or differences in how stellar masses are measured in simulations and observations.}
   {}


\titlerunning{Towards a consistent framework of comparing galaxy mergers in observations and simulations}

\authorrunning{Wang et al.}

\maketitle

\section{Introduction}

In the standard lambda cold dark matter (LCDM) cosmology framework, structure formation is hierarchical in nature, which means dark matter halos and galaxies that reside within them grow through successive accretion and mergers over time (White \& Rees 1978; Cole et al. 2000; Fakhouri \& Ma 2008; Wetzel et al. 2009), although there is no simple mapping between halo mergers and galaxy mergers. The potential impact of galaxy mergers in the context of galaxy formation and evolution is wide ranging, including the assembly of stellar mass (Guo \& White 2008;  Bundy et al. 2009; Bridge et al. 2010; Robaina et al. 2010; Rodriguez-Gomez et al. 2016; Mundy et al. 2017; Qu et al. 2017; Duncan et al. 2019), the triggering of intense starburst activity and accretion onto the central supermassive black hole (De Lucia \& Blaizot 2007; Di Matteo et al. 2012; Ellison et al. 2013; Torrey et al. 2014; Pearson et al. 2019b; Patton et al. 2020; McAlpine et al. 2020; Gao et al. 2020), the change in the chemical content in galaxies (e.g. Kewley, Geller \& Barton 2006; Ellison et al. 2008; Torrey et al. 2012; Moreno et al. 2015; Bustamante et al. 2018; Thorp et al. 2019), and the transformation of morphological, structural, and kinematic properties (Robertson et al. 2006a,b; Naab \& Burkert 2003; Burkert et al. 2008; Villalobos \& Helmi 2008; Purcell et al. 2009; Hopkins et al. 2010; Dubois et al. 2016; Rodriguez-Gomez et al. 2017). 

However, we still lack a detailed quantitative understanding of the role of mergers in galaxy evolution (e.g. how mergers drive star formation and nuclear activity across cosmic history) compared to other physical processes (such as smooth accretion and violent disc instability; e.g. Ceverino, Dekel \& Bournaud 2010; Cacciato, Dekel \& Genel 2012; Porter et al. 2014). Two main stumbling blocks are the difficulty in detecting statistically large, reliable, and representative merger samples across a wide range in cosmic history and the difficulty in comparing observations and theoretical models in a consistent and meaningful way. 
  
As galaxy mergers have diverse, complex, and sometimes very subtle features, detecting mergers has always been a challenging task. There are two main families of identification techniques, one which selects close galaxy pairs in 3D space and one which selects morphologically disturbed galaxies due to either ongoing or very recent merger activity. The close pair method requires highly complete spectroscopic samples and so  generating large samples is observationally very expensive (Lin et al. 2004 or Woods \& Geller 2007). Moreover, a significant fraction could be contaminating flybys, which are close pairs that are interacting with each other but will detach from each other at a later time (Sinha \& Holley-Bockelmann et al. 2012; Lang et al. 2014). The relative frequency of flyby events versus mergers could also change as a function of redshift and dark matter halo mass. The morphological indicator method requires high-resolution and high-quality imaging to detect merging signs (such as tidal tails, strong asymmetry, double nuclei). Quantitative morphological statistics, such as the CAS parameters (concentration, asymmetry, clumpiness) of Conselice (2003) and Fourier-mode asymmetry measurements (Peng et al. 2010), the Gini coefficient  (Abraham, van den Bergh \& Nair 2003), and $M_{20}$ index (Lotz et al. 2004), suffer from poor accuracy and low completeness (Huertas-Company et al. 2015; Snyder et al. 2019). One can also use a qualitative approach to selecting morphologically disturbed galaxies through visual inspection as human brains can easily identify complex patterns. The Galaxy Zoo (GZ) project has engaged citizen scientists  through an online platform to visually classify around two million galaxies since its launch over a decade ago (Lintott et al. 2011). However, even GZ is not scalable to the amount of data emerging from current and upcoming surveys detecting many hundreds of millions to billions of galaxies. In addition, human visual classification is subjective and not easily reproducible. Using ideal mock images of paired galaxies from the Next Generation Illustris Simulations (IllustrisTNG), Blumenthal et al. (2020) identified distinct biases in merger samples constructed from visual identification. 

More recently, new methods for detecting mergers have been developed. Still in the category of looking for signs of merging activity in the imaging data, there have been several studies that employ deep convolutional neural networks (deep CNNs) to detect mergers (Ackerman et al. 2018; Pearson et al. 2019a,b; Bottrell et al. 2019; {\'C}iprijanovi{\'c} et al. 2020; Ferreira et al. 2020). This technique is able to reproduce visual classification in a fraction of the time it would take humans and does not require any hands-on engineering of features in order to solve a given classification or regression problem (feature learning). Deep CNNs are also capable of capturing the full range of diversity in the appearances of mergers and are very robust to some of the challenging characteristics of the imaging data (such as noise, resolution, and artefacts). However, this technique is fundamentally limited by the quality of the training data used to optimise the free parameters in the deep CNNs. Beyond relying solely on imaging data, it is also possible to include kinematic information to aid the identification of mergers with the availability and growing popularity of integral field spectroscopy (IFS; Shapiro et al. 2008; Mason et al. 2017; Simons et al. 2019). However, again this technique is observationally very expensive and so difficult to generate large merger samples across a wide range in redshift and galaxy properties.

On the other hand, even when we do have a reasonably good merger sample identified from observations (based on morphological signs or physically close pairs), it is often extremely hard to determine its true reliability and completeness against parameters such as gas content, orbital parameters, orientation, stellar mass, mass ratio, and redshift, which then prohibits detailed and quantitative comparisons between different observational studies as well as with cosmological simulations in order to derive constraints on theoretical models of galaxy formation and evolution (e.g. Berrier et al. 2006; Genel et al. 2009; Williams, Quadri \& Franx 2011; Moreno et al. 2013). In other words, we currently lack a consistent framework within which we can make like-for-like comparisons between mergers in observations and mergers in simulations in order to place not only qualitative but also quantitative constraints on theoretical models. 

In recent years,  it has become possible to make large-volume, cosmological, hydrodynamical simulations of galaxy formation and evolution with statistically significant and increasingly more realistic galaxy populations, using for example the EAGLE  (Evolution and Assembly of GaLaxies and their Environments) simulation suite (Crain et al. 2015; Schaye et al. 2015), the Illustris simulations (Vogelsberger et al. 2014a, b; Genel et al. 2014), Simba (Dav{\'e} et al. 2019), or IllustrisTNG (Springel et al. 2018; Pillepich et al. 2018b). These simulations, which employ a comprehensive galaxy formation model and state-of-the-art numerical code, are successful in reproducing a wide range of observations including the cosmic star-formation history, stellar population properties, stellar mass functions, scaling relations, clustering properties, galaxy sizes, and morphologies (Furlong et al. 2015; Sparre et al. 2015; Nelson et al. 2018; Pillepich et al. 2018a; Springel et al. 2018; Rodriguez-Gomez et al. 2019). The close interaction between these detailed simulations and observations is key to continuously improving our understanding of the complex physics involved, as simulations are crucial for properly interpreting observational results which in turn provide constraints and feedback into the physical models used in the simulations, thus enabling further fine tuning and iterations. The powerful synergy between theory and observations can only be fully achieved through quantitative and consistent comparisons between the two. Among other things, this requires us to forward model the simulation data to produce fully realistic mock observations that include observational effects such as background, noise, angular resolution, chance superpositions, and redshift dimming.

In this paper we set out the first steps towards building such a consistent framework. First, we carry out forward modelling of ideal synthetic images of simulated galaxies from the IllustrisTNG simulations in order to include major observational effects. We then apply the same merger-detection method, based on deep CNNs (building on our previous work in Pearson et al. 2019a,b), to both the simulated mock observations and real observations. Finally, we carefully examine and compare the fraction of major mergers as a function of stellar mass in the observations and simulations.  Characterising the dependence of the major merger fraction on stellar mass provides important information that can be used in theoretical models of galaxy formation and evolution. Significant differences between the trends in observations and in simulations will highlight problems with models of galaxy formation and evolution.

This paper is organised as follows. In Section 2, we describe the relevant data products from the Galaxy And Mass Assembly (GAMA) survey, the Kilo-Degree Survey (KiDS), and the IllustrisTNG simulations.  In Section 3, we explain our forward-modelling approach to include major observational effects in turning ideal synthetic images of simulated galaxies into realistic mock observations. We also briefly describe our merger-detection method, which is based on deep CNNs. In Section 4, we present our results on the dependence of the major merger fraction on stellar mass in the low-redshift Universe and compare with previous measurements. Finally, we give conclusions in Section 5.

\section{Data}

In this section, we introduce the main characteristics of the observational datasets and the simulation dataset used in this paper. 

\subsection{The Galaxy And Mass Assembly (GAMA) survey and the Kilo-Degree Survey (KiDS)}

GAMA\footnote{\url{http://www.gama-survey.org}} is an optical spectroscopic survey of low-redshift galaxies, mainly conducted at the Anglo-Australian Telescope (Driver et al. 2009, 2011; Liske et al. 2015). GAMA covers three equal-sized fields (each of which spans $\sim5\times12$ degrees in size) to an apparent SDSS DR7 (Sloan Digital Sky Survey - Data Release 7) Petrosian r-band magnitude limit of $r = 19.8$ mag at $>98\%$ completeness: G09, G12, and G15 (centred at a right ascension of $\sim$9, 12, and 14.5 hours, respectively) on the celestial equator. We use spectroscopic redshifts from GAMA and limit our sample to a thin redshift slice $0.1<z<0.2$,  which results in a sample of 16950, 21716, and 22599 galaxies in G09, G12, and G15, respectively. We use stellar mass estimates from the GAMA survey (Wright et al. 2017) estimated using the MAGPHYS (da Cunha et al. 2008; da Cunha \& Charlot 2011)  spectral energy distribution (SED)  fitting tool and 21 band photometric information ranging from the far-ultraviolet to the far-infrared (Wright et al. 2016). Full details of the MAGPHYS fits can be found in Driver et. al. (2016). Due to the optical selection limit  of the GAMA survey $r<19.8$ mag, the galaxy stellar mass limit for a volume-complete sample out to $z=0.2$ is $\sim10^{10.3}M_{\sun}$ (Wright, Driver \& Robotham 2018). As a result, the observations are affected by incompleteness effects at the low-mass end $<10^{10.3}M_{\sun}$. The number of galaxies above the stellar mass completeness limit is 7488, 8969, and 9255 in G09, G12, and G15, respectively.

For the purpose of detecting major mergers using deep CNNs, we need to produce cut-out images of each GAMA galaxy in our sample. We choose the KiDS survey which overlaps with GAMA in the three equatorial fields. KiDS\footnote{\url{http://kids.strw.leidenuniv.nl}} (de Jong et al. 2013) is a high spatial resolution and high sensitivity ESO public optical survey carried out with the VLT Survey Telescope (VST) and OmegaCAM camera, which will image 1350 deg$^2$ in four filters (u, g, r, i) in single epochs per filter, with mean seeing in the r-band of around  0.7\arcsec (roughly a factor of two better than SDSS) and mean limiting magnitudes (5$\sigma$) of 24.23, 25.12, 25.02, and 23.68 in u, g, r, and i, respectively. We use the latest data release 4 (DR4; Kuijken et al. 2019).

\subsection{The IllustrisTNG simulations}

Building upon the original Illustris project (Vogelsberger et al. 2014a,b; Genel et al. 2014), the IllustrisTNG simulations (Springel et al. 2018; Pillepich et al. 2018b; Nelson et al. 2018; Marinacci et al. 2018; Naiman et al. 2018) are a suite of cosmological magneto-hydrodynamical simulations that model a range of physical processes considered relevant for the formation of galaxies, for example star formation and evolution, gas cooling, chemical enrichment, and stellar and supermassive black hole (SMBH) feedback, carried out using the moving mesh AREPO code (Springel 2010).  Apart from the observables (e.g. cosmic star-formation history, the local galaxy stellar mass function, and galaxy sizes at $z=0$) which the IllustrisTNG model was tuned to match (Pillepich et al. 2018a), the model has also been shown to successfully reproduce several key observables across different redshifts, such as galaxy morphology and structure, galaxy colour bimodality, and spatial clustering properties.

We choose the highest resolution version, `TNG100', which follows the evolution of 2 $\times$ $1820^3$ resolution elements within a periodic cube of $75h^{-1}$ Mpc per side,  corresponding to an average mass of the baryonic resolution elements of $1.39\times10^{6}M_{\sun}$. The spatial scale of the simulation is $0.5h^{-1}$ kpc, which is essentially set by the gravitational softening length of the dark matter and stellar particles. Rodriguez-Gomez et al. (2019) generated synthetic images of galaxies from the IllustrisTNG and the original Illustris hydrodynamic cosmological simulations using the SKIRT radiative transfer code (Baes et al. 2011; Camps \& Baes 2015), including the effects of dust attenuation and scattering. However, these images were generated without including any companion galaxies. Although it would be straightforward to include the companions from the same friends-of-friends (FoF) group, running SKIRT with full dust radiative transfer is computationally very expensive. Therefore, instead we created images without taking the dust into account (which is 100 to 1000 times faster). The differences in morphology are fairly small (Rodriguez-Gomez et al. 2019). Bottrell et al. (2019) also showed that adding the effect of dust seems to make very little difference to classification performance of the deep CNNs.

We choose the redshift snapshot number 87, corresponding to redshift $z=0.15$. The only criterion we used for generating images of the simulated TNG galaxies is that the stellar mass, defined as the total stellar mass of the galaxy as measured by the SUBFIND halo-finding algorithm (Springel et al. 2001), should be higher than $10^9 M_{\sun}$. Rodriguez-Gomez et al. (2019) showed that TNG galaxies with stellar mass $> 10^{9.5} M_{\odot}$ at low redshifts have reliable morphologies. In the present study, in order to increase the sample size with which to train the CNN classifier, we lowered the stellar mass limit to $10^9 M_{\odot}$. Therefore, simulated galaxies at the lower mass end between $10^9 M_{\odot}$ and $10^{9.5} M_{\odot}$ may be less well resolved which might have an effect on the relative number of false positives and false negatives over this mass range. In Section 4 when we compare the stellar mass dependence of the merger fraction in observations and simulations, we use a different stellar mass definition, which is  the stellar mass within twice the stellar half-mass radius, in order to exclude the intracluster light  from massive galaxies. Images are created centred at the galaxy of interest and from three different viewing angles (projections along the x, y, and z axes in the simulated volume), taking into account that the galaxy is located at $z=0.15$. The three viewing angles are essentially random, because there are no chance alignments between the galaxies and the axes of the box. The images are made to be 128x128 pixels with the pixel scale set to  0.21 arcsec/pixel. For full details on how the synthetic images are generated, please refer to Rodriguez-Gomez et al. (2019).

In order to train deep CNNs to recognise major mergers and non-mergers, we need to select the two categories from TNG100. First, we define mergers as merging galaxies that will eventually merge within the next 1 Gyr (i.e. pre-mergers) or merged galaxies for which the merging event took place within the previous 500 Myr (i.e. post-mergers). We then require that the mass ratio (which is defined as the ratio of the most massive primary galaxy to the least massive secondary galaxy) be $<4:1$ as the mass ratio of the merging progenitor galaxies is known to have a major effect on the appearance of the mergers. Furthermore, the mass ratio is based on the stellar masses of the two merging galaxies at the time when the secondary reached its maximum stellar mass (Rodriguez-Gomez et al., 2015). In total, we obtain 1019 major mergers from the snapshot number 87 from TNG100, which is 4.6\% of the total number of galaxies with stellar mass $>10^{9}M_{\odot}$ in that snapshot. 

We did carry out some tests using a wider range in mass ratio but the performance (e.g. accuracy) of the deep CNNs degrades quite significantly. We also tried to use a smaller range in merging history (e.g. by selecting merging galaxies that merge within the next 300 Myr or merged galaxies for which the merging event happened within the previous 300 Myr). However, the resulting sample size is too small to train the CNNs. Similarly, dividing the major mergers into pre-mergers and post-mergers resulted in smaller samples and poorer performance of the CNNs.

To construct the non-merger sample, we first define a minor merger sample by selecting all mergers with the range  $4:1<$ mass ratio $<10:1$ that will merge within the next 1 Gyr or that merged within the previous 500 Myr. Galaxies in the TNG100 that are not classified as a major or minor merger according our definition are then classified as non-mergers.

\section{Method}

In this section, we describe our forward-modelling procedure designed to simulate the major observational effects and our merger identification method based on deep CNNs.

\begin{figure}
\includegraphics[height=7.in,width=3.in]{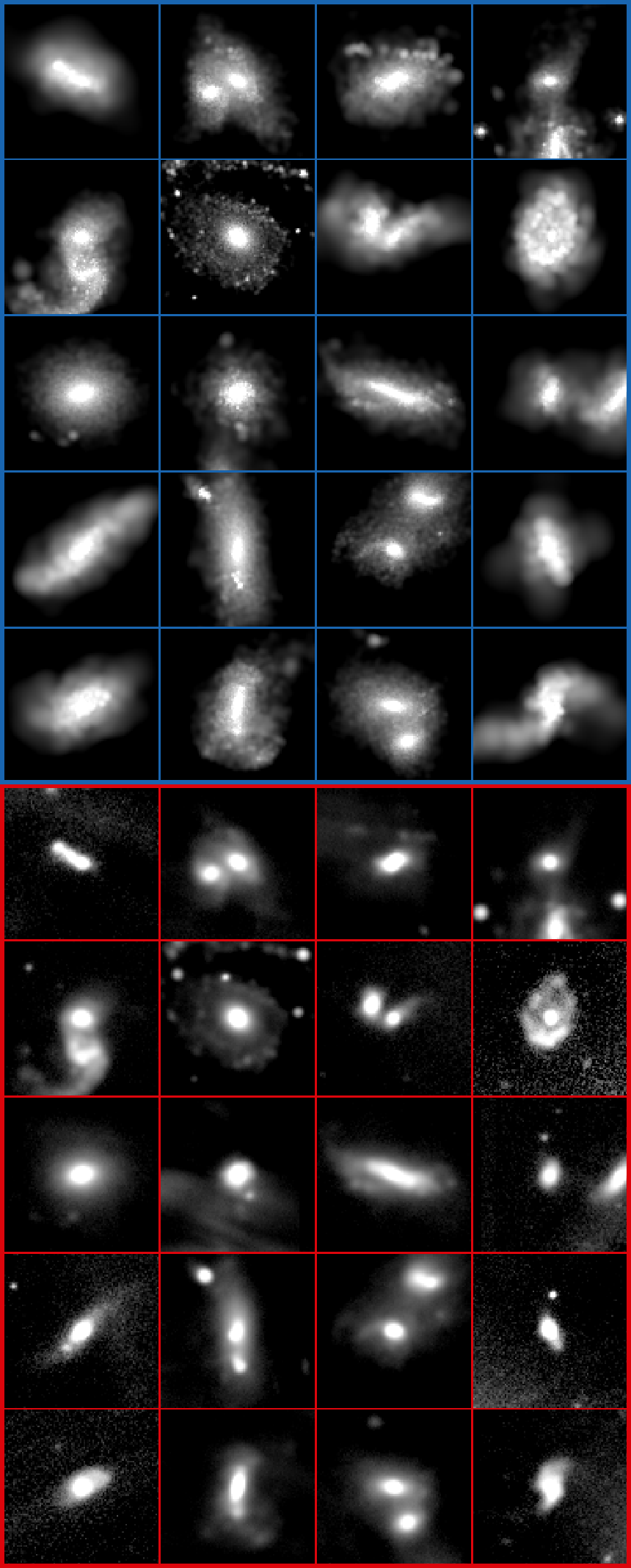}
\caption{Randomly selected TNG major mergers before (top five rows outlined in blue) and after adding observational effects (bottom five rows outlined in red). Major mergers in the pre-merging phase are easier to identify by eye than mergers in the post-merging phase. Merging features such as tidal tails become much more difficult to identify once observational effects are included.}
\label{tngmerger}
\end{figure}

\begin{figure}
\includegraphics[height=7.in,width=3.in]{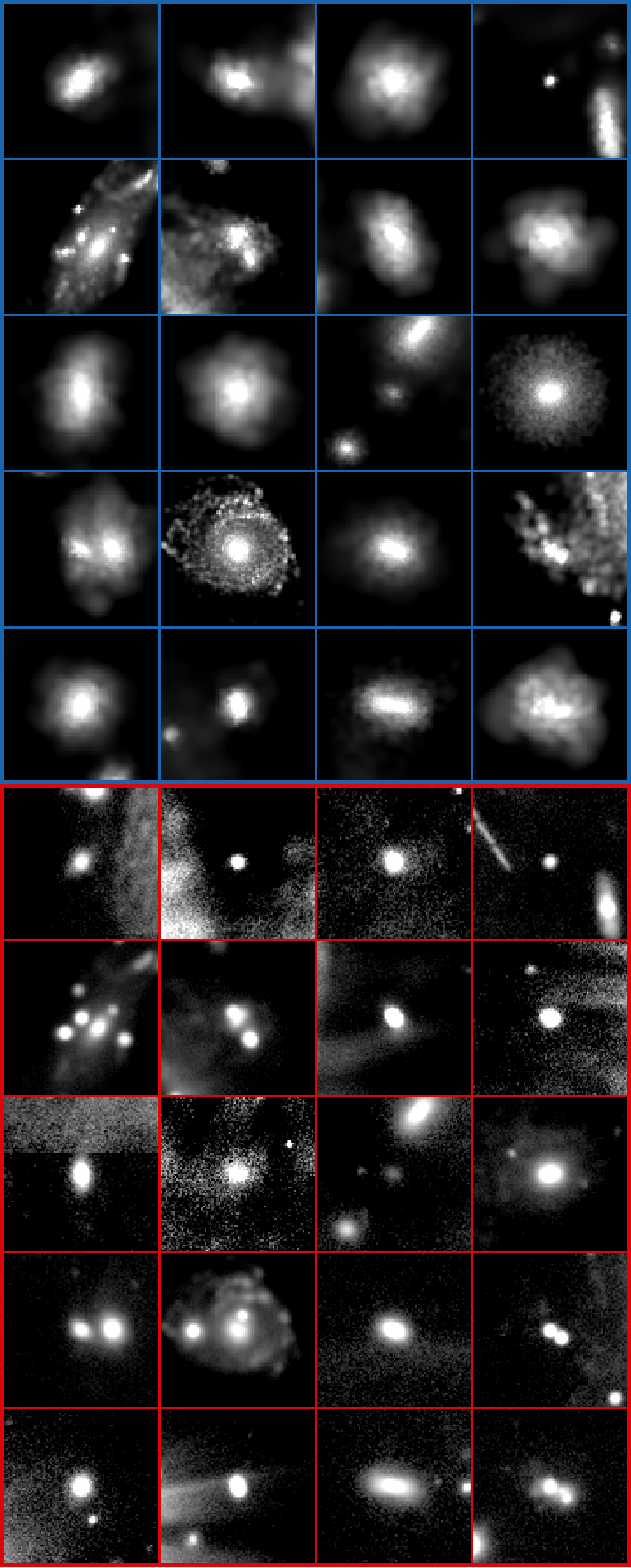}
\caption{Randomly selected TNG non-mergers before (top five rows outlined in blue) and after adding observational effects (bottom five rows outlined in red). Chance projections with objects in the real KiDS images could cause the classification method to incorrectly identify a non-merger as a merger.}
\label{tngnonmerger}
\end{figure}

\subsection{Making mock observations using forward modelling}

To make ideal synthetic IllustrisTNG images look similar to those from the KiDS survey, the simulated images were created with a pixel scale of 0.21 arcsec/pixel, the same as KiDS, and were then convolved with a synthetic, Gaussian point spread function with a full width at half maximum (FWHM) of 0.77 arcsec, which corresponds to the average FWHM of KiDS. The resulting images were then injected into real KiDS observations, avoiding overlaps with real galaxies, to add realistic image noise, sky background, artefacts, and chance projections. To do this injection, we generated 1\,000\,000 random coordinates within each KiDS tile, removing any positions that are within 128 arcsec of any sources in the KiDS catalogue. For the positions that remain, a 256$\times$256 pixel cut-out is made from the KiDS image, centred on the RA-Dec of the randomly generated position and ensuring there are data inside the entire cut-out. This cut-out will contain real, observed noise as well as chance projections from undetected sources. For each IllustrisTNG galaxy, an empty KiDS cut-out is randomly selected and the cut-out is added to the IllustrisTNG image, ensuring the units of the IllustrisTNG images match those of the KiDS noise. This random selection is done with replacement so more than one IllustrisTNG galaxy can use the same noise cut-out.

To scale the images, we determine the 25th percentile of the linearly scaled image. The image is then log scaled to bring out the low-surface-brightness features (Bottrell et al. 2019) and we determine the median of the logged flux density for the pixels that were not below the 25th percentile of the linear image. The pixels found to be below this 25th percentile were then set to the median logged flux density. The log-scaled image was then clipped at the 99th percentile and normalised to a maximum of unity.

Figures~\ref{tngmerger} and~\ref{tngnonmerger} show 20 randomly selected examples of TNG major mergers and non-mergers before and after adding observational effects. In the raw images before observational effects are added, major mergers in the pre-merging phase show clear merging features such as tidal tails and flow of material between the two merging galaxies. Major mergers in the post-merging phase in comparison are more difficult to identify. We expect that more of the post-mergers will be classified as non-mergers because of a lack of conspicuous merging features. It can be clearly seen that the merging features become significantly less recognisable after including observational effects as low surface brightness features tend to be much weaker or lost.  For the major mergers and non-mergers, there are also examples of chance projections with objects in the real KiDS images. These chance projections could cause the CNNs to mis-classify a non-merger into a major merger.

\subsection{Major merger identification with deep convolutional neural networks}

In the last decade, great strides have been made in image classification and computer vision, using CNNs. The CNN, based on modelling complex patterns with a hierarchy of increasingly abstract representations, is one of the best-performing deep learning network architectures. A common architecture for CNNs consists of several convolutional and pooling layers followed by one or more fully connected layers before the output. The convolutional layers which apply trainable convolution kernels to an input image in order to make feature maps (or activation map) are the most important building block of a CNN. Each neuron in a convolutional layer is connected only to neurons located within a small rectangular receptive field in the previous layer. In comparison, fully connected layers are the same as layers in a classic artificial neural network with fully connected architecture, i.e. each neuron from one layer is connected to all of the neurons in the previous layer. The pooling layers are used to subsample the input image to speed up training and limit the risk of overfitting.  These layers split the inputs from the previous layer into small rectangles and reduce these rectangles to a single value. This value is often either the average of the values in the rectangle or the maximum, as is the case here. The output layer of a CNN is usually a single vector of class scores.

This hierarchical structure (in which higher-level features are obtained through the successive assembly of those at lower levels) is common in real-world images, which is one of the main reasons why CNNs work so well for image-recognition tasks. Thanks to graphics processing units (GPUs) and large training sets, CNNs have been shown to equal or even exceed human accuracy on some complex tasks of visual perception (LeCun et al. 2015). Once trained, CNNs can perform reproducible visual-like classifications in a fraction of the time it would take humans. There are many other advantages of deep CNNs. For example, CNNs do not require pre-designated/hand-crafted features (known as feature engineering), and are very versatile (although not entirely domain-independent). Leading software companies are already using CNNs to revolutionise real-life applications (such as self-driving cars and automatic classification systems for images and videos). Dieleman et al. (2015) was one of the first to use CNNs in galaxy morphology classification and achieved >99\% accuracy for images, with high agreement among Galaxy Zoo citizen scientist classifiers. Since then, deep CNNs have become increasingly more popular in astronomy (e.g. Huertas-Company et al. 2015; Hoyle 2016; Petrillo et al. 2017; Kim \& Brunner 2017; Davies et al. 2019; Canameras R., et al., 2020), often offering striking  improvement compared to conventional methods or other types of machine learning methods for a wide range of astronomical problems (e.g. identifying lensed galaxies, estimating photometric redshifts, classifying morphological types, and detecting transients).

\begin{figure}
\includegraphics[height=2.6in,width=3.54in]{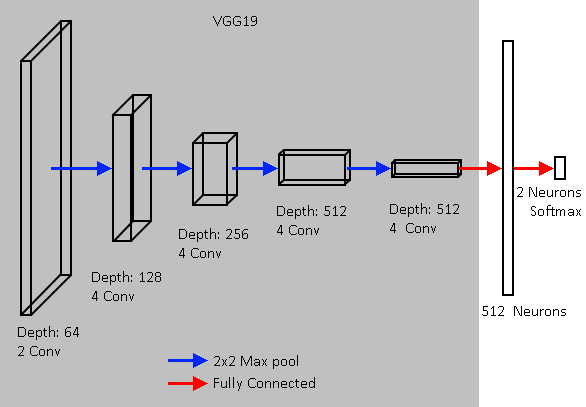}
\caption{Architecture of the CNN used to identify major mergers and non-mergers. The pre-trained VGG network (shaded area) is used with the top three fully connected layers removed. There are five blocks containing 2, 4, 4, 4, and 4 convolutional layers with 64, 128, 256, 512, and 512 kernels respectively. All kernels are $3\times3$ pixels and each block is connected with a $2\times2$ pixel max pooling layer (blue arrows). We add a single fully connected layer with 512 neurons and a fully connected output layer with 2 neurons (unshaded area). The input to the network is a $88\times88$ pixel image with three identical channels.}
\label{tng_network}
\end{figure}

\begin{figure}
\includegraphics[height=2.3in,width=3.in]{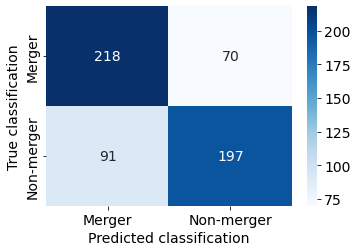}
\caption{Confusion matrix for TNG images classified by the network, showing 218 true positives (TP), 91 false positives (FP), 197 true negatives (TN), and 70 false negatives (FN).}
\label{confusion}
\end{figure}

\begin{figure}
\includegraphics[height=7.in,width=3.in]{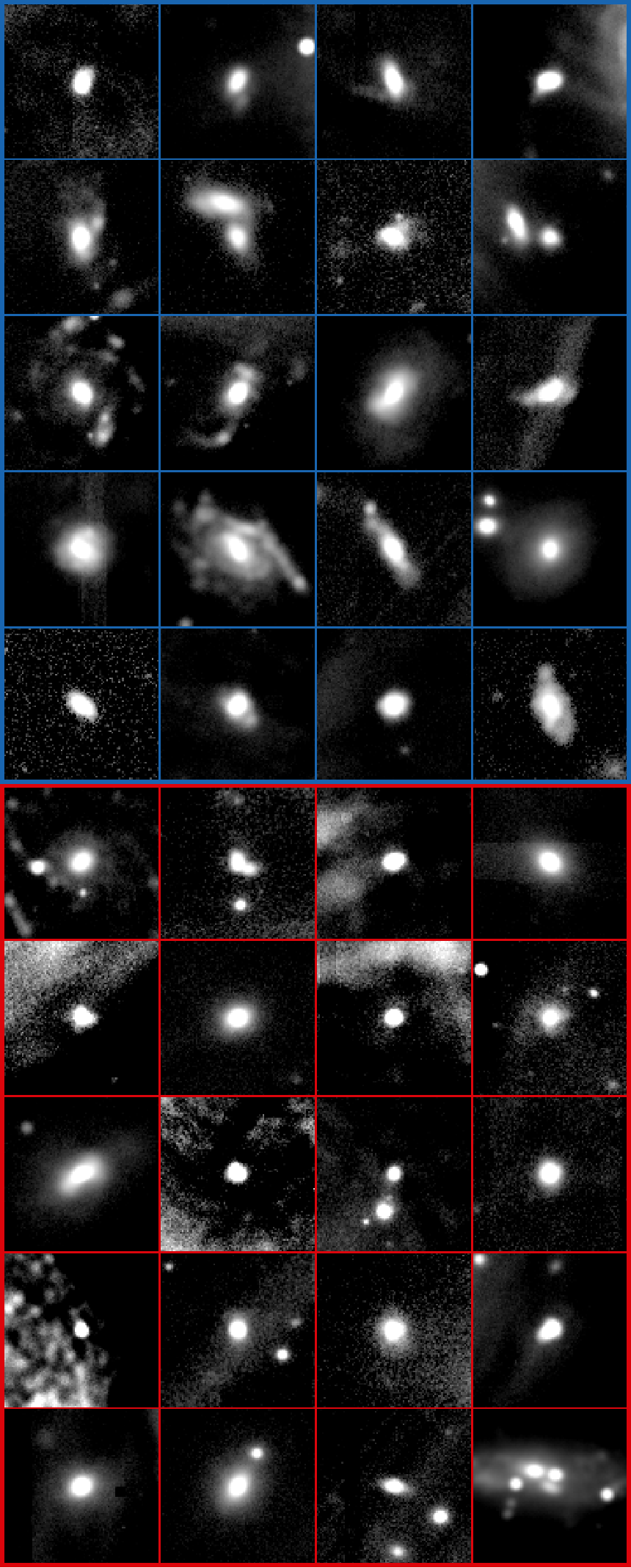}
\caption{Randomly selected examples of true TNG major mergers that are correctly and incorrectly identified by our network. The top five rows (outlined in blue) show examples of true major mergers that are correctly identified as major mergers. The bottom five rows (outlined in red) show examples of true major mergers that are incorrectly identified as non-mergers. Major mergers that are correctly identified by the network show more conspicuous merging signs (such as more obvious distortions and/or the presence of a merging companion.)}
\label{TNG-TP-FN}
\end{figure}

\begin{figure}
\includegraphics[height=7.in,width=3.in]{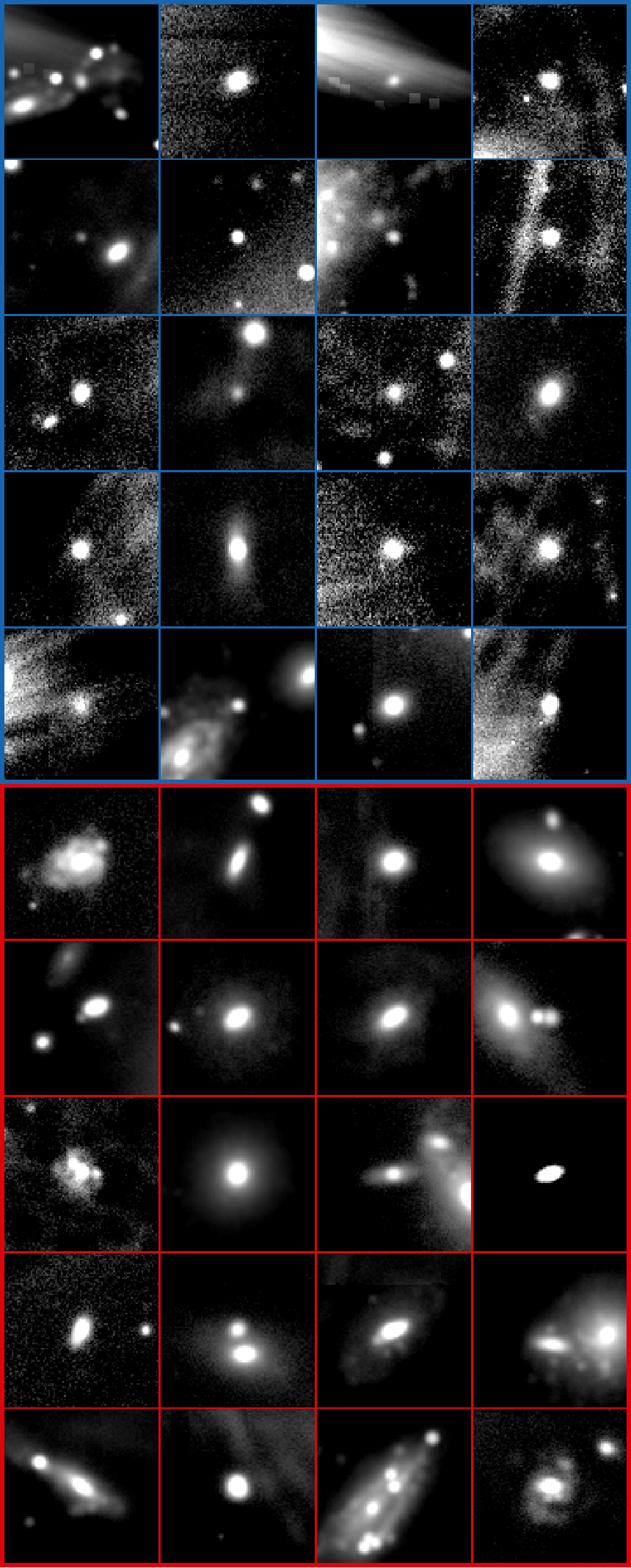}
\caption{Randomly selected examples of true TNG non-mergers that are correctly and incorrectly identified by our network. The top five rows (outlined in blue) show examples of true non-mergers that are correctly identified as non-mergers. The bottom five rows (outlined in red) show examples of true non-mergers that are incorrectly identified as major mergers.}
\label{TNG-TN-FP}
\end{figure}

\begin{figure}
\includegraphics[height=7.in,width=3.in]{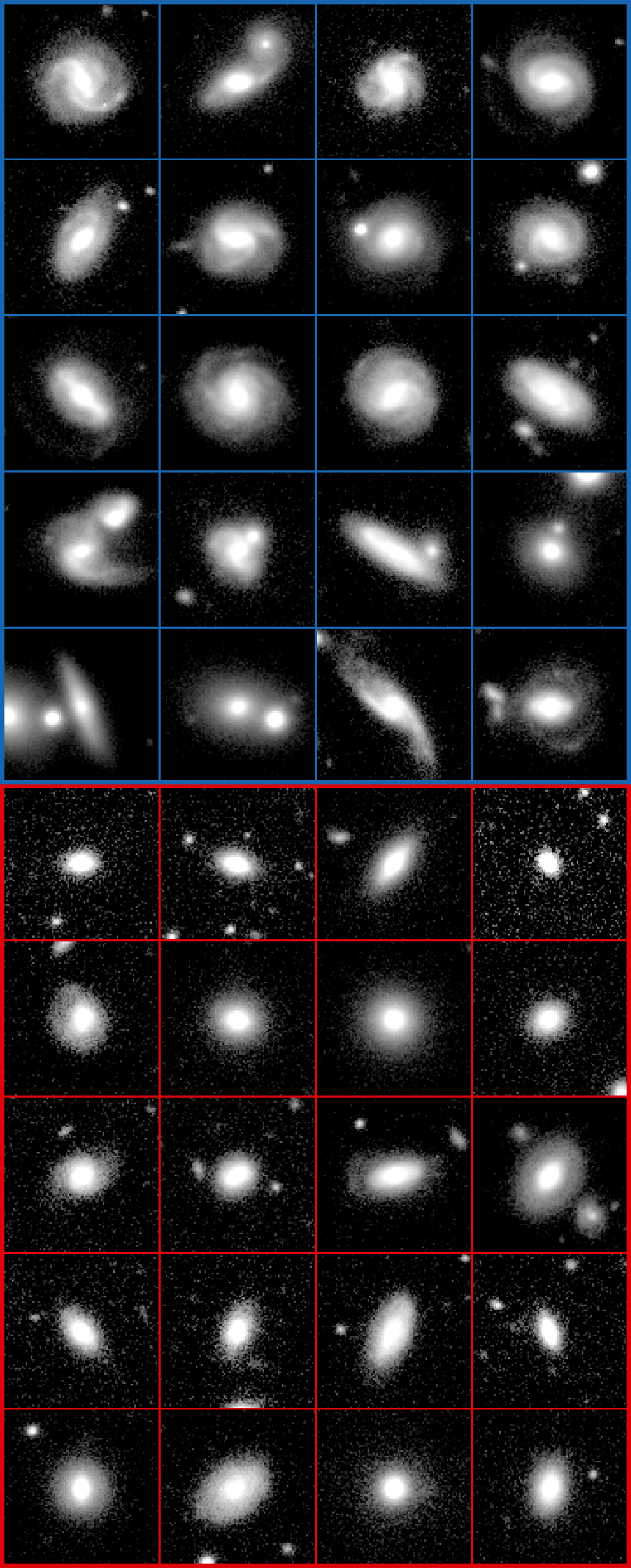}
\caption{Randomly selected KiDS galaxies that are identified as major mergers (top five rows outlined in blue) and non-mergers (bottom five rows outlined in red) by our network. Non-mergers are relatively featureless and isolated.}
\label{KiDS-GAMA09}
\end{figure}

To identify merging galaxies, we perform transfer learning using the pre-trained VGG19 convolutional neural network \citep[CNN;][]{2014arXiv1409.1556S} within the Keras framework \citep{chollet2015keras} utilising the TensorFlow \citep{tensorflow2015-whitepaper} backend. In the first application of deep CNNs for the purpose of detecting galaxy mergers (which are relatively rare astronomical objects), Ackermann et al. (2018) demonstrated the power of transfer learning for small datasets. Transfer learning works by pre-training on millions of natural images (such as cars, dogs, and cats) in order to learn rich feature representations for a wide range of images. As a result, transfer learning is able to improve the generalisation power of the trained classifier, thereby leading to a better overall classification performance for small datasets. We chose the VGG19 neural network model as it is one of the state-of-the-art deep-learning image classifiers and is included in the Keras deep-learning library. VGG19 has also been used with astronomical data before in the classification of compact star clusters in nearby spiral galaxies (Wei et al. 2020).

The VGG19 neural network has been pre-trained with the ImageNet data set \citep{deng2009}, a collection of several million images of everyday objects  (animals, keyboards, pencils, mugs, etc.), and has had the top three fully connected layers removed. We add a further two layers: a fully connected layer, with 512 neurons and rectified linear unit\footnote{A rectified linear unit (ReLU) computes a linear function of the inputs and outputs the result if it is positive, and zero otherwise. It is mainly used to increase the non-linearity of the activation function.} \citep{nair2010rectified} activation, and an output layer with 2 neurons, that is, one neuron for the major merger  classification and one for
the non-merger classification,  and softmax activation. These classifications provide the probability for each class in the range of [0, 1] that sum to unity. In the following, we use the output for the major merger class \texttt{frac\_merger}, which is equivalent to using the output for the non-merger class as it is 1 - \texttt{frac\_merger} in our binary classification. When training the network, only these two additional layers are trained using the Adam optimiser \citep{2014arXiv1412.6980K} while the weights and biases in the VGG19 network are left unaltered. Figure~\ref{tng_network} illustrates the architecture of the CNN used in this paper to detect major mergers and non-mergers.

To train the network, we use images from the IllustrisTNG cosmological, hydrodynamical simulation.  The IllustrisTNG data set comprises 1019 major mergers that are within 1 Gyr before or 500 Myr after the merger event and a further 1019 non-merging galaxies (so that the two classes are balanced in size), all at a simulation redshift of 0.15. For each galaxy, we take three r-band images in different orientations, resulting in 6114 unique galaxy images. These images are then split into three groups: 80\% of the images are used to train the network, 10\% are used to measure the progress of the training (validate the network), and the final 10\% are used to test the network. To increase the size of the training sample and reduce the rotational dependance inherent to CNNs, when training we also randomly rotate the images by multiples of $10^{\circ}$ when they are loaded, creating an effective training size of 88\,560 galaxies per classification. This is not done for the validation or testing samples. The input images are then cropped to 88$\times$88 pixels and stacked with themselves to create three identical colour channels for the images; VGG19 requires input images with three colour channels. Different bands were not used for the images due to storage limitations when creating the KiDS cutouts.

Once trained, we follow Pearson et al. (2019a) and alter the threshold between major mergers and non-mergers so that a threshold where fall-out and recall are closest to (0,1) for the validation data set is used; here this is 0.53. With this cut between major mergers and non-mergers, we find an accuracy of 0.720, recall of 0.757, precision of 0.706, specificity of 0.684, and negative predictive value of 0.738. In this paper, major mergers are designated as
the positive class (`P') and non-mergers as the negative class (`N'). Therefore, true positives are major mergers classified as major mergers and false positives are non-mergers classified as major mergers. Similarly, true negatives are non-mergers classified as non-mergers and false negatives are major mergers classified as non-mergers. Fall-out is the fraction of false positives (FP) with respect to the sum of true positives (TP) and false negatives (FN), i.e. FP / (TP + FN). Recall is defined as TP / (TP+FN). Precision is defined as TP / (TP + FP). Specificity is the fraction of true negatives (TN) with respect to the total number of TN and FP, i.e. TN/(TN+FP). Negative predictive value is defined as TN/(TN+FN). Accuracy is defined as (TP+TN)/(TP+FP+TN+FN). Figure~\ref{confusion} shows the confusion matrix for the TNG images of mergers and non-mergers classified by the network. The confusion matrix is a useful tool for evaluating the performance of the classifier. Each element in the confusion matrix represents the number of times objects of class A are classified as class B. Our classifier correctly identifies 218 major mergers and 197 non-mergers, however it incorrectly identifies 91 non-mergers as major mergers and 70 major mergers as non-mergers.

As discussed in Pearson et al. (2019a), the lower accuracy of a simulation-trained network relative to a network trained using visually identified (merger and non-merger) samples is a result of differences in the training samples. Visually identified samples by construction only contain mergers with conspicuous merging features and therefore can be highly incomplete (e.g. Blumenthal et al. 2020). Simulations have the advantage that we have the ground truth with which to tune the network. By construction, training samples selected from the simulations are both reliable and complete. However, many true mergers (especially post-mergers) could lack obvious merging features (see Fig. 1 and Fig. 4). On the other hand, the performance of our network trained with TNG samples seems to be better than our previous network trained with galaxies from the EAGLE simulation, which achieved an accuracy of 65.2\%. This could be due to differences in merger definition, differences in the CNN architectures, and/or differences in galaxy morphologies in the different simulations. It is also worth pointing out that Bottrell et al. (2019) achieved better performance applying deep CNNs to identify mergers in the Feedback In Realistic Environments (FIRE) simulations (Hopkins et al. 2018). However, there are many important differences between our study and that of Bottrell et al. (2019). For example, the training sample of Bottrell et al. (2019) is constructed from a suite of non-cosmological binary galaxy interaction simulations. Also, to boost sample size, the binary galaxy interaction simulations are finely sampled in time which means there could be a high level of correlation among the merger samples.

Figure~\ref{TNG-TP-FN} shows 20 randomly selected examples of true TNG major mergers that are correctly and incorrectly identified by our network. The main cause of misclassification (i.e. classifying true major mergers as non-mergers) appears to be the lack of conspicuous merging features in the true major mergers. Furthermore, most of them seem to be in the post-merging phase or the merging companion is behind the primary galaxy. Figure~\ref{TNG-TN-FP} shows 20 randomly selected examples of detected TNG non-mergers that are correctly and incorrectly identified.  The main cause of misclassification (i.e. classifying true non-mergers as major mergers) seems to be the presence of chance projections or clumpy structures in the images.  In Appendix A, we further investigate   which parts of an image the CNN classifier is most sensitive to in making a classification of merger versus non-merger through a series of occlusion experiments.

The trained deep CNNs using training samples from the TNG simulations are then applied to the real KiDS images of 55\,627 GAMA galaxies with redshifts between 0.1 and 0.2. Figure~\ref{KiDS-GAMA09} shows 20 randomly selected KiDS images of GAMA galaxies that are identified as major mergers and non-mergers by our network. Galaxies that are classified as major mergers exhibit a range of merging morphologies. In some cases, the merging companion can also be identified. Galaxies classified as non-mergers appear relatively featureless (more smooth and round compared to galaxies classified as major mergers) and in some cases isolated.

\section{Results}

In this section, we present our results after applying the trained deep CNNs to the test dataset from the TNG simulations and the real GAMA/KIDS data. It is important to note that we are only using the test set  when discussing the TNG major mergers in this section because the network has been tuned using the training set and the validation set (see Section 3.2).

\begin{figure}
\includegraphics[height=2.8in,width=3.84in]{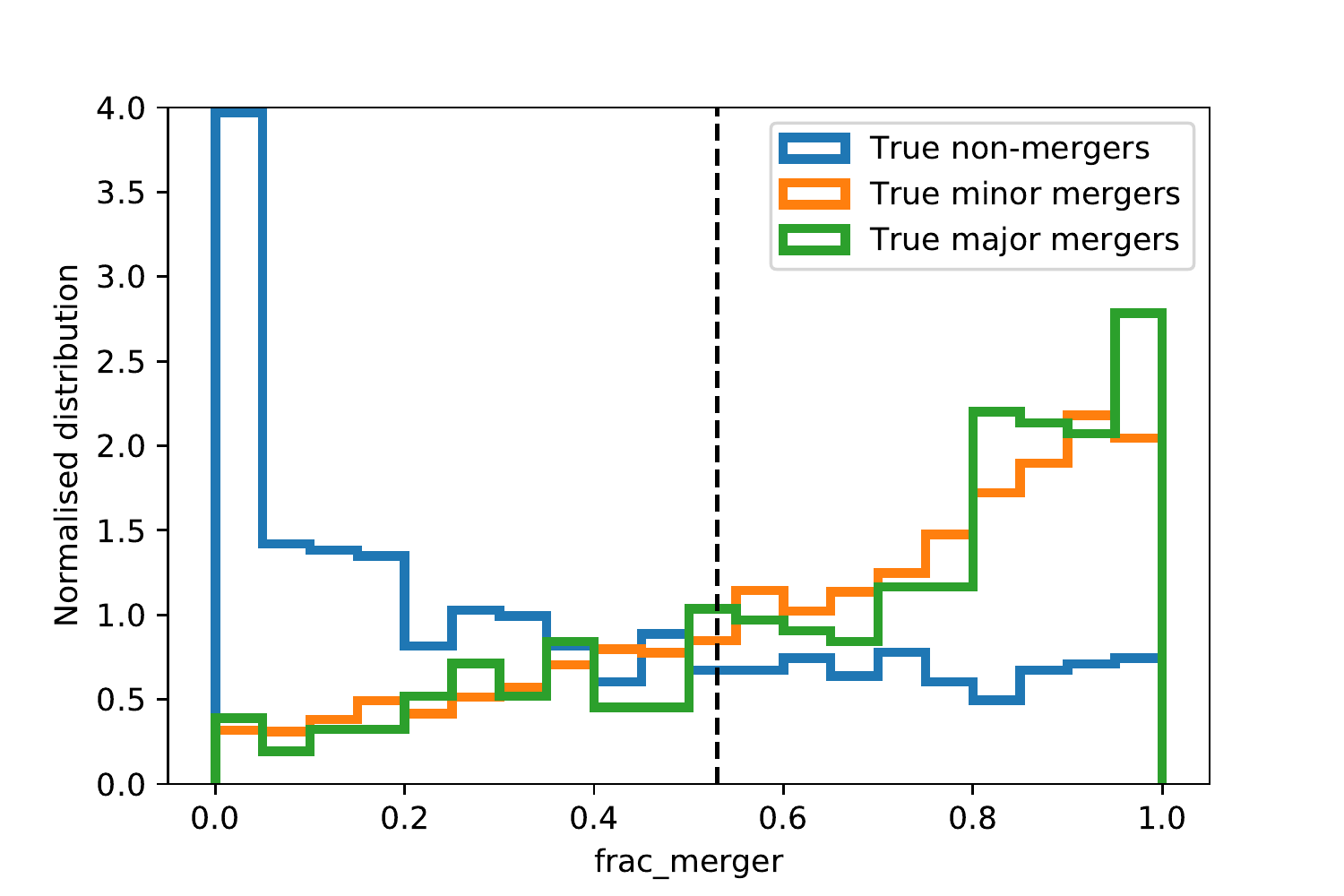}
\caption{Normalised distribution (meaning the area under each histogram sums to unity) in the CNN-derived probability of major merger for the real major mergers, real minor mergers, and real non-mergers in the test set from the TNG simulations. The vertical dashed line indicates the threshold of  \texttt{frac\_merger} $>0.53$, which is used to define a major merger detected by the network.}
\label{frac}
\end{figure}

\begin{figure}
\includegraphics[height=2.8in,width=3.84in]{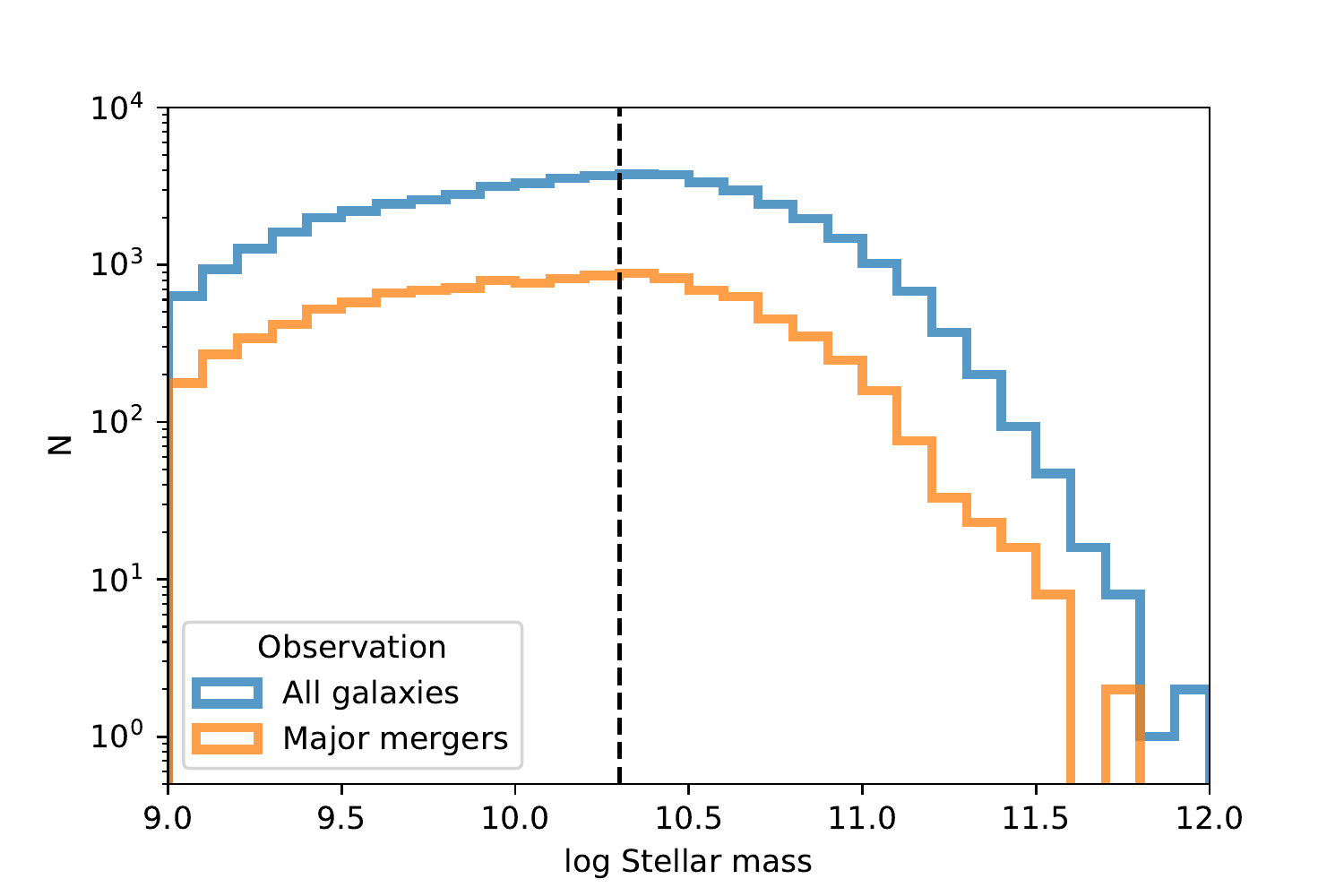}
\caption{Number of all galaxies and detected major mergers as a function of stellar mass in the GAMA/KiDS observations. The vertical dashed line indicates the mass complete limit (galaxy stellar mass $=10^{10.3}M_\odot$) of a volume-complete sample out to $z=0.2$. Our sample is clearly affected by incompleteness below $10^{10.3}M_\odot$.}
\label{mass_dep}
\end{figure}

Classifying all images from the test set with our CNN merger detection classifier gives us a probability of between 0 and 1 for every test set image. A value of zero indicates classification as a non-merger galaxy, and a value of 1 indicates classification as a major merger system.  Figure~\ref{frac} compares the normalised distribution of the CNN-derived probability of being a major merger for the true major mergers, true minor mergers, and  the true non-mergers in the test set from the TNG simulation. As discussed in Section 3.2, we use a threshold of \texttt{frac\_merger} $>0.53$ to define a major merger. Reassuringly, the peaks of these probability distributions are located in the right places, with major mergers above 0.53 and non-mergers below. However, there is a tail in the probability distribution for the true major mergers at \texttt{frac\_merger} $<0.53$ and there is a tail in the probability distribution for the true non-mergers at  \texttt{frac\_merger} $>0.53$. The prominent peak at \texttt{frac\_merger} $\sim0$ for the non-mergers is due to the fact that isolated galaxies without disturbed morphologies can be classified as non-mergers with high confidence. We also apply the trained CNN to true minor mergers in the TNG simulation as the real dataset will include major mergers, minor mergers and non-mergers. It seems that true minor mergers are more likely to be classified as major mergers than non-mergers by our network. The contamination of minor mergers is present in both the observational datasets and the simulations. This should not have a big impact on our results, as we are only interested in the relative differences between the observations and simulations in the shape of the major merger fraction as a function of stellar mass.

Figure~\ref{mass_dep} shows the number of all galaxies and major mergers detected by the trained network as a function of stellar mass in the GAMA/KIDS observations. As described in Section 2.1, the observations are affected by incompleteness effects at the low-mass end, that is $<10^{10.3}M_{\sun}$. The detected major merger fraction is 22\%, 23\%, and 24\% in G09, G15, and G12, respectively. Our major merger fraction is higher compared to other studies in the literature. For example, Casteels (2014) identified highly asymmetric galaxies as merger candidates and then determined which ones are major mergers (by examining close pairs) to obtain major merger fractions of $<5\%$ for nearby galaxies. Man et al. (2016) found a major merger fraction of around 5\% over the redshift range $0.1<z<0.4$ using close pairs selected from the $K_s$-band selected UltraVISTA catalogue (Muzzin et al. 2013). This is expected given the accuracy of the CNNs. Some genuine major mergers are missed while some non-major-mergers (including minor mergers and non-mergers) will be classified as major mergers. Given the relatively small fraction of major mergers, a lot more non-major-mergers will be classified as major mergers than the other way around. However, major mergers are relatively rare in both real observations and simulations and so the contaminations will affect the observations and simulations in a similar way. In this paper, we are mainly interested in the differences and similarities between the observations and simulations in the relative merger fraction as a function of stellar mass. Another cause of the discrepancy with previous results in the literature could be the fact that major merger samples identified in previous studies are highly incomplete as they tend to include only mergers with obvious merging features.

 \begin{figure}
\includegraphics[height=2.8in,width=3.84in]{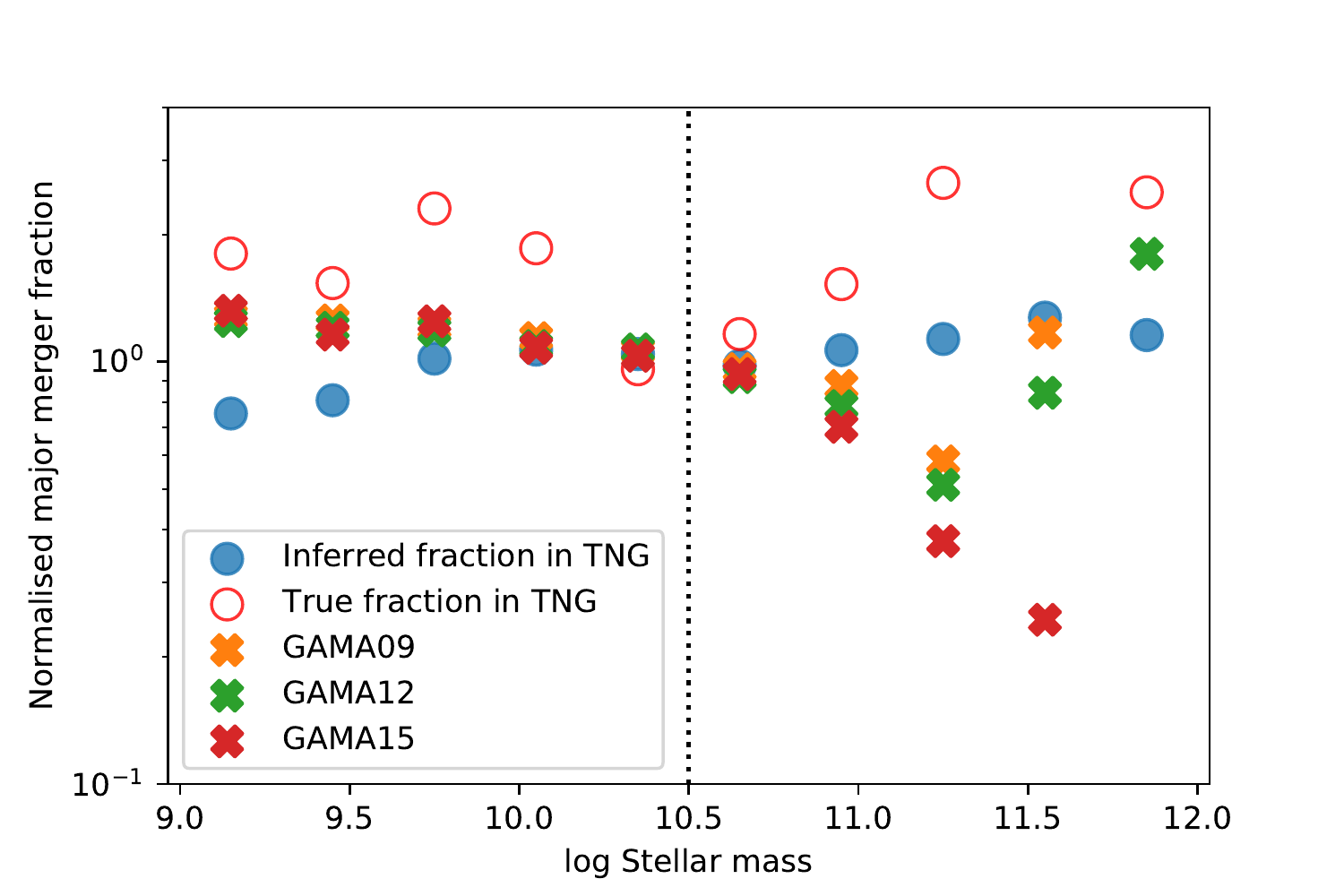}
\caption{Dependence of the major merger fraction on galaxy stellar mass in the GAMA/KiDS observations and in the TNG simulations. The trends in the three GAMA fields agree well with each other, except at the most massive end which is severely limited by sample size. The vertical dotted line marks where galaxy stellar mass $=10^{10.5}M_{\odot}$ , which is where kinetic AGN feedback takes over at late times in the TNG simulations.}
\label{mass_ratio}
\end{figure}

There have been many previous attempts to measure the mass dependence of the merger fraction and merger rate using studies of close pairs or morphologically disturbed galaxies. However, there is still no consensus on the mass dependence. Some studies have found evidence for constant or a slight increasing trend with stellar mass (e.g. Xu et al. 2004; Patton \& Atfield 2008; Domingue et al. 2009; Xu et al. 2012; Besla et al. 2018), while others have found a decreasing trend (Bridge, Carlberg, \& Sullivan 2010). Specifically, Casteels et al. (2014) found that the mass-dependent major merger fraction is fairly constant in the range $10^{9.5} < M < 10^{11.5}M_{\odot}$ and increases with decreasing stellar mass to about 4\% at $M_*\sim10^{9}M_{\odot}$. In Fig.~\ref{mass_ratio}, we compare the dependence of the major merger fraction on stellar mass in observations and in the test set from the TNG simulation. We plot the normalised major merger fractions by scaling the fractions at galaxy stellar mass $=10^{10.5}M_{\odot}$ to unity. This is because our TNG samples are designed to be class-balanced (i.e. the relative fractions of each class have been adjusted) as mentioned in Section 3.2. The inferred stellar mass dependence of the major merger fraction (which is derived as the fraction of detected major mergers over the total number of galaxies) in the TNG simulation is very smooth. In comparison, the true major merger fraction (derived as the fraction of true major mergers over the total number of galaxies) in the TNG simulation varies significantly with stellar mass, with a dip around galaxy stellar mass $\sim10^{10.5} M_{\odot}$. Interestingly, this is also the characteristic stellar mass of galaxy colour transition in the TNG model which is set by supermassive blackhole feedback in its low-accretion state (Weinberger et al. 2018; Nelson et al. 2019; Donnari et al. 2019; Terrazas et al. 2020).

The difference between the inferred and true mass dependence of the major merger fraction in TNG is not surprising given the accuracy of the classifier. However, given that the main purpose of this study is to investigate the differences and similarities between observations and simulations, a more informative comparison is that between the inferred mass dependence of the major merger fraction in TNG and the observed mass dependence of the major merger fraction from the GAMA/KiDS observations. As we apply the same merger-detection method to the two datasets, in principle we have the same problems with systematic effects such as misclassification and incompleteness. The trends in the three GAMA fields (G09, G12 and G15) agree very well with each other, except at the most massive end (at $\log (M_*/M_{\odot})> 11.5$) where we are severely limited by small number statistics (see Fig. 8). The dependence of the major merger fraction on stellar mass has a fairly mild decreasing trend with increasing stellar mass  at $9<\log (M_*/M_{\odot})<11.5$ in the GAMA/KiDS observations. Over the mass range $9.5<\log (M_*/M_{\odot})<10.5$, the inferred mass dependence of the major merger fraction in TNG is very similar to the trend in the observations. Below stellar mass $\sim10^{9.5} M_{\odot}$, the simulated galaxies in TNG have less reliable morphologies (Section 2.2) and therefore caution is needed when interpreting the trends at the low-mass end. Above galaxy stellar mass $\sim10^{10.5} M_{\odot}$, the trends are very different as the inferred mass dependence in TNG remains flat while the mass dependence in the observations decreases rapidly with increasing stellar mass (albeit with large dispersion among the three GAMA fields at stellar mass $\gtrsim 10^{11.5} M_{\odot}$).

We hypothesise that  the difference in the inferred major merger fraction as a function of stellar mass between the observations and the TNG simulations may be related to the SMBH feedback model in the simulations. It is well known that SMBH feedback plays an important role in setting the relation between the stellar mass content and host dark matter halo mass for more massive halos (e.g. Croton et al. 2006; Somerville et al. 2008; Silk \& Mamon 2012; Pillepich et al. 2018a). Additionally, the merger fraction depends on the host dark matter halo mass (Khochfar \& Burkert 2001; Hopkins et al. 2008). Therefore, any changes in the mapping between stellar mass and halo mass at the high-mass end due to the specific prescriptions of the SMBH feedback could have a direct impact on the shape of the major merger fraction as a function of stellar mass. However, there could be other, more mundane reasons behind the differences seen at the high-mass end. For example, the volume of the TNG100 simulation is around a factor of three smaller compared to the volumes probed by each of the three GAMA equatorial fields. Given the spread at the high-mass end among the three GAMA fields, it is reasonable to expect that the measured data points of the simulations will be subject to similar or even larger uncertainty. Finally, the differences in how stellar masses are measured in the observations and simulations could also play a role (Pillepich et al. 2018b).

\section{Conclusion}
 
This is the first study towards building a consistent framework to meaningfully compare the stellar mass dependence of the major merger fractions in observations and cosmological simulations using deep-learning techniques. We first perform forward modelling of the ideal synthetic images of simulated galaxies in IllustrisTNG to include major observational effects such as angular resolution, noise, sky background, redshift dimming, and chance projections. We then train deep CNNs to discriminate between major mergers and non-mergers in the simulations, achieving an accuracy of 72\%. Finally we apply the trained network with the optimised parameters to detect major mergers in the KiDS images of GAMA galaxies in the redshift range $0.1<z<0.2$. 

Thanks to this set of procedures, we are now able to compare observations and simulations in a consistent way. We find that the dependence of major merger fraction on stellar mass over the mass range $[10^{9.5}M_{\odot}, 10^{10.5}M_{\odot}]$ is smooth, and is similar between the GAMA/KiDS observations and the IllustrisTNG simulation. Above a stellar mass of $\sim10^{10.5}M_{\odot}$, the mass dependence trends in the simulations and observations are different. This could be linked with SMBH feedback in its low-accretion state in the TNG simulations, which causes a sharp transition in the median colour from blue to red at a characteristic mass of $10^{10.5}M_{\odot}$. The differences at the high-mass end between the simulations and observations could also be due to the relatively small volume of the simulations and/or the differences in how stellar masses are measured. In the GAMA/KiDS observations, the major merger fraction decreases in a fairly mild fashion from stellar mass $M_*\sim10^9 M_{\sun}$ to $M_*\sim10^{11} M_{\sun}$, by a factor of approximately two. 

There are also some differences in the inferred and true major merger fraction in the TNG simulations as a function of stellar mass, which is a consequence of the performance of the CNN-based merger identification method. To further improve the performance (such as accuracy, precision, and recall) of the  network, we need much larger training samples, which could be obtained by combining different redshift snapshots and/or combining different simulations. In future work, we will extend our investigations by characterising the dependence of major merger fraction on redshift, environment (characterised by dark matter halo mass), and distance from the galaxy star-formation main sequence (which can be used as an indicator of starburst intensity). We will also extend our binary merger classification to include merger stages such as pre- and post-mergers.

\begin{appendix}

\section{Occlusion heat maps}

One drawback with deep CNN is that its internal representation of the problem is obscure and not easily interpretable. To visualise and understand how CNN makes a classification, we perform a series of occlusion experiments (Zeiler \& Fergus 2013; Pearson et al. 2019a) to extract the most important features within the input images that the network uses to identify a merging or non-merging galaxy. A $8\times8$ pixel area (or patch) within the images that were correctly identified by the network is made black by setting the RGB values to zero. We then slide this region (i.e. the occluder) across the image in steps of one pixel in both x and y directions, generating images with different $8\times8$ pixel areas masked. For each masked image, the output for the merger class is recorded as a function of occluder position. To determine which regions of the image have the greatest effect on the classification, we generate a heat map for each object where each pixel is the average of the merger class outputs of the images where that pixel is masked.

\begin{figure*}
\includegraphics[height=1.in,width=3.6in]{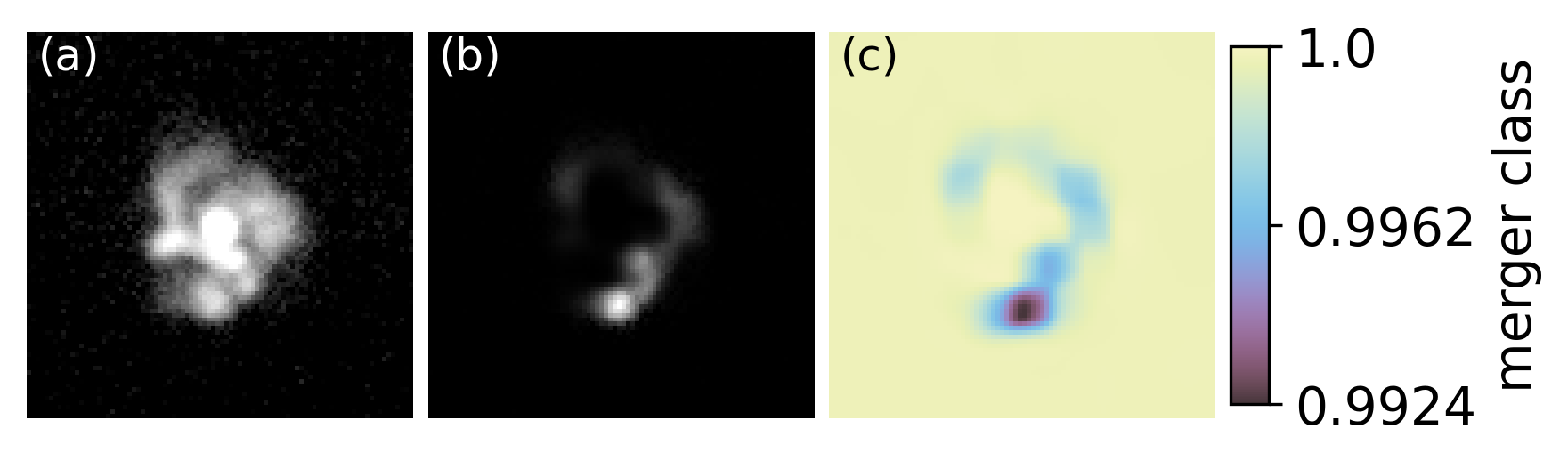}
\includegraphics[height=1.in,width=3.6in]{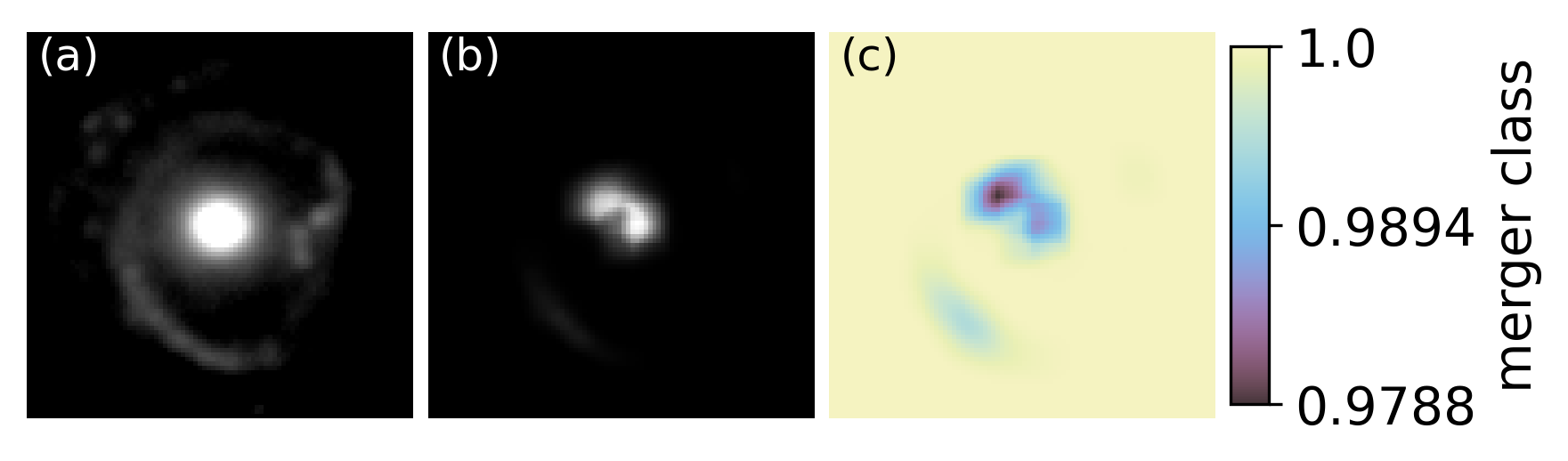}
\includegraphics[height=1.in,width=3.6in]{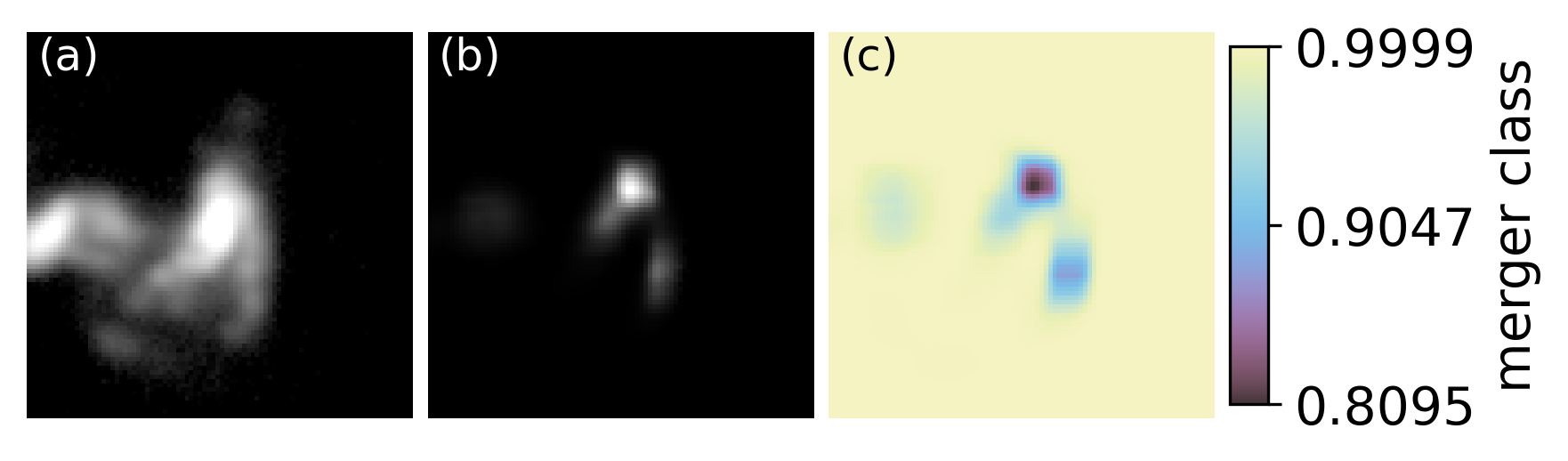}
\includegraphics[height=1.in,width=3.6in]{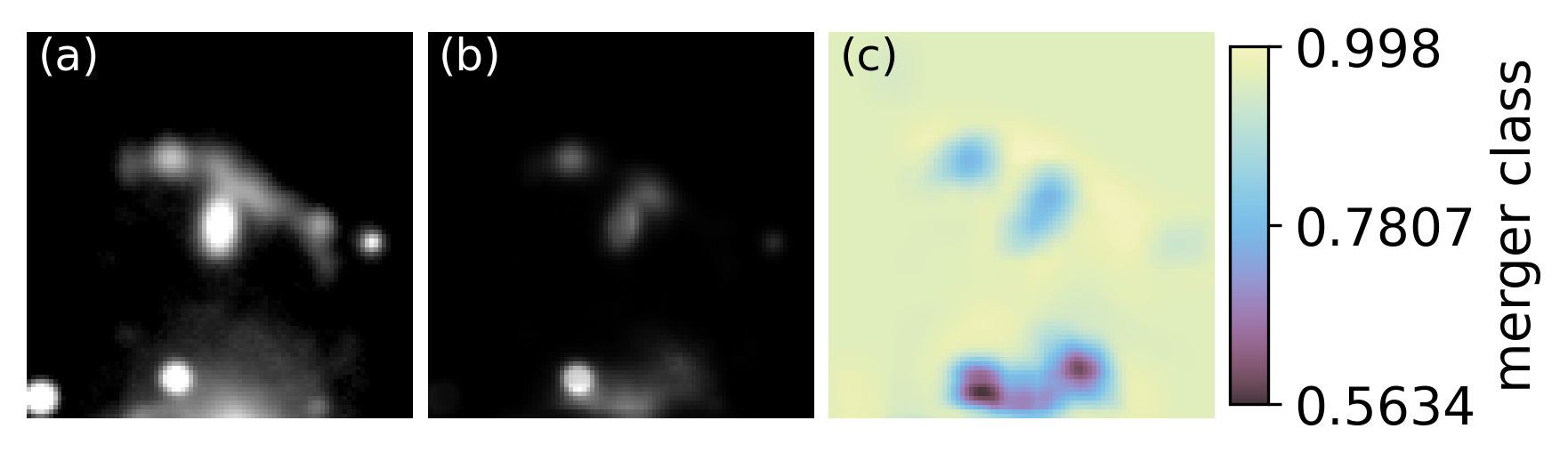}
\includegraphics[height=1.in,width=3.6in]{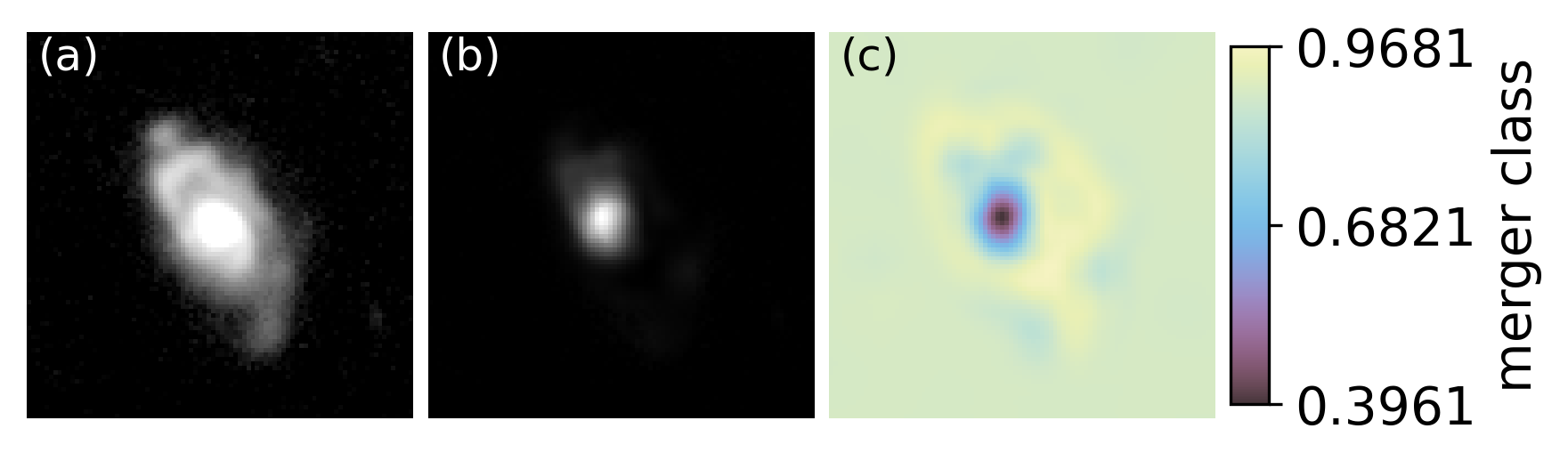}
\includegraphics[height=1.in,width=3.6in]{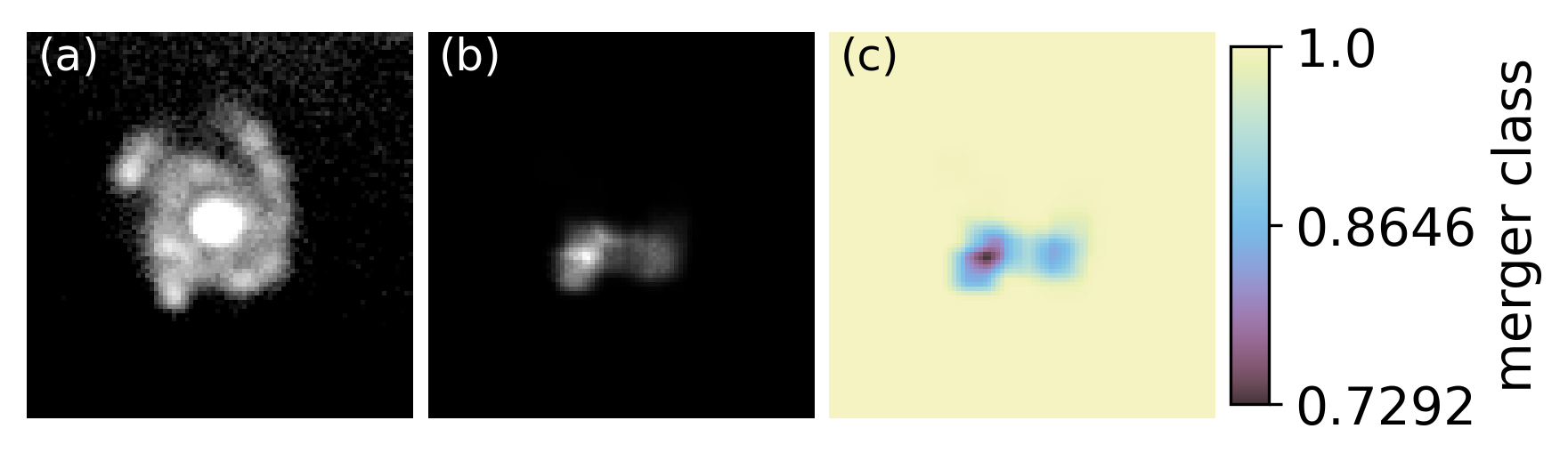}
\caption{Heat maps demonstrating how the network detect example major mergers. Panels (a) show the original image of the merging galaxy being classified, panels (b) show the regions that most effect the merger classification and panels (c) show the heat maps where regions with darker colours have a greater effect on the classification.}
\label{heatmap_merger}
\end{figure*}

\begin{figure*}
\includegraphics[height=1.1in,width=3.6in]{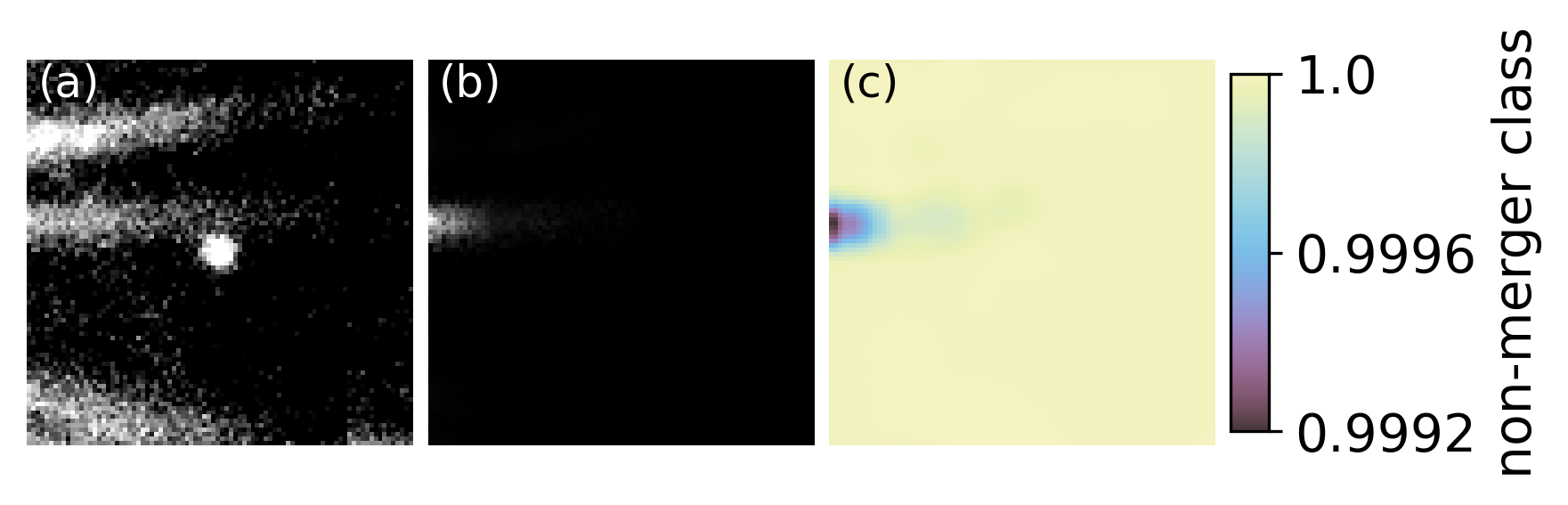}
\includegraphics[height=1.1in,width=3.6in]{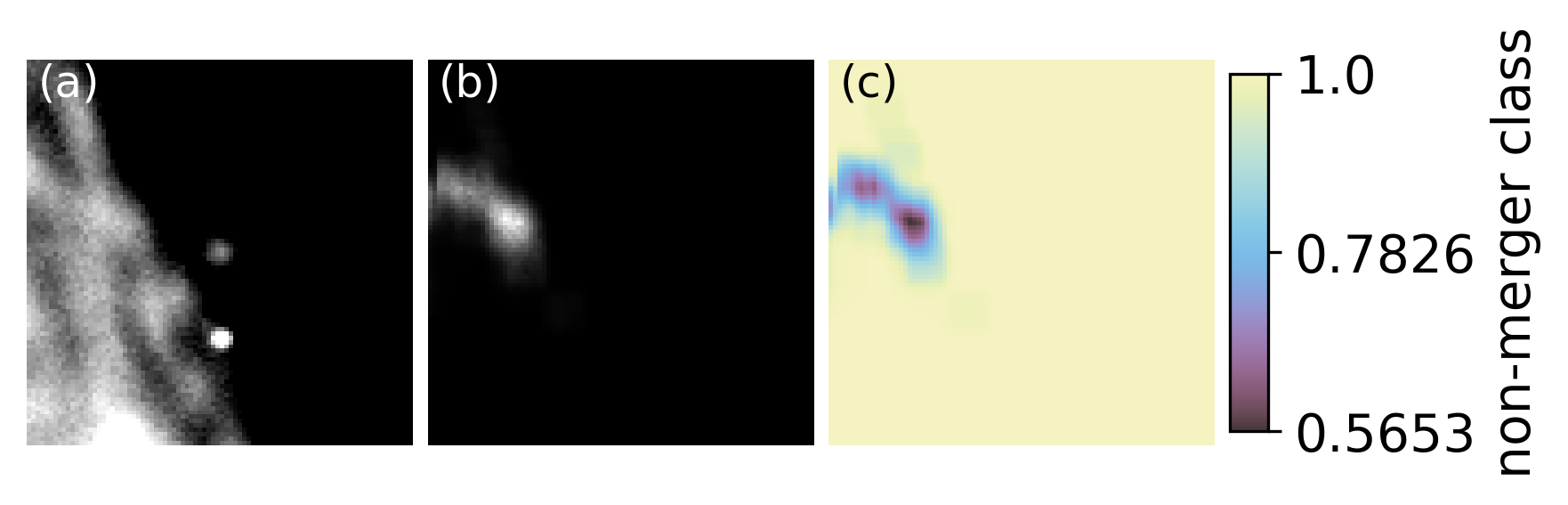}
\includegraphics[height=1.1in,width=3.6in]{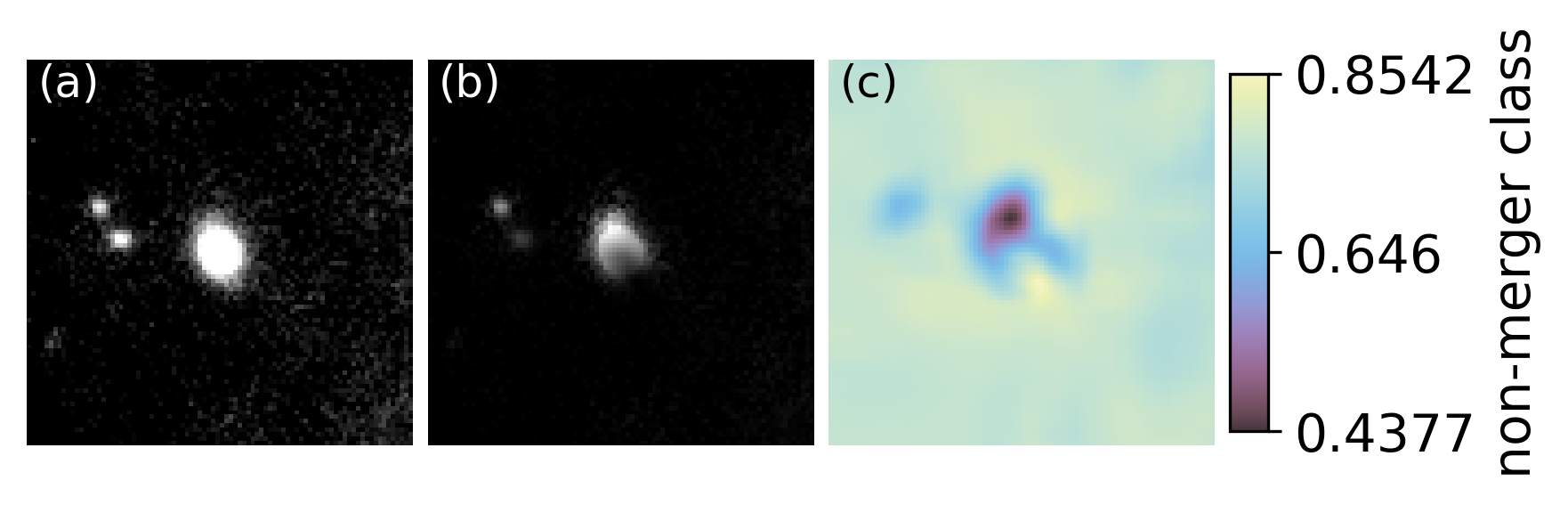}
\includegraphics[height=1.1in,width=3.6in]{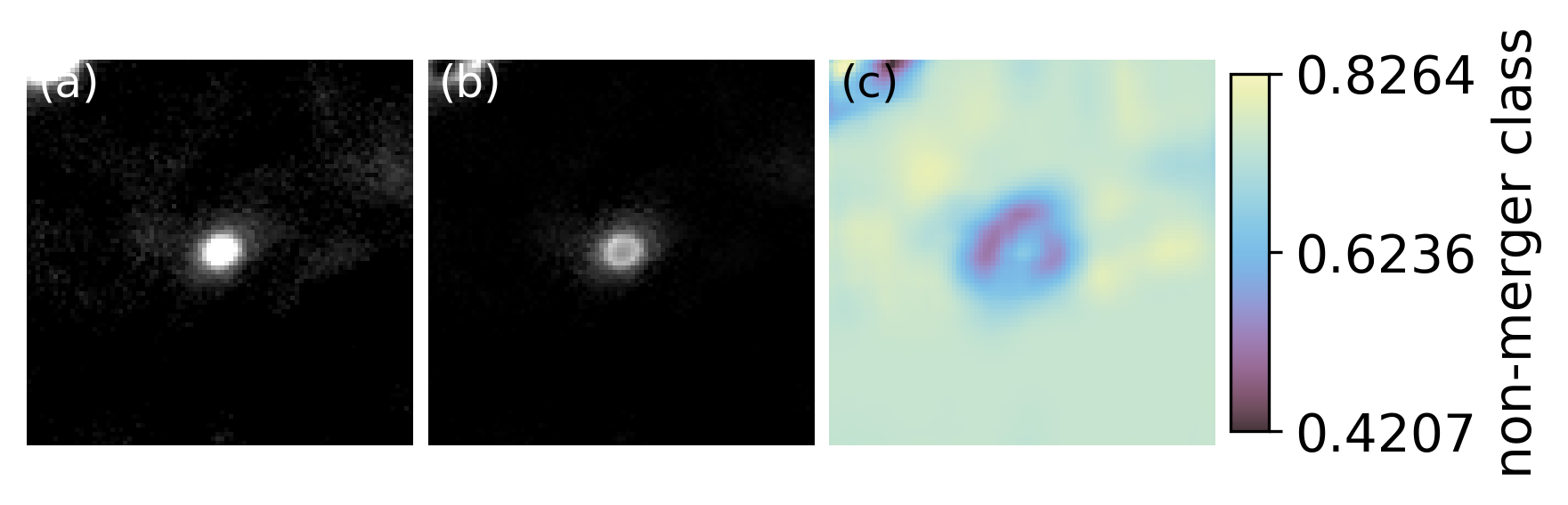}
\caption{Heat maps demonstrating how the network detect example non-merging galaxies. Panels (a) show the original image of the non-merging galaxy being classified, panels (b) show the regions that most effect the merger classification and panels (c) show the heat maps where regions with darker colours have a greater effect on the classification.}
\label{heatmap_nmerger}
\end{figure*}

Figure ~\ref{heatmap_merger} shows examples of major mergers galaxies that were correctly identified by the CNN. Panels (a) in Fig. ~\ref{heatmap_merger} show the original image being classified. Panels (b) show the regions that have the greatest influence on the merger classification. Panels (c) show the heat maps where regions with darker colours have a greater effect on the classification (lower merger class output). Panels (b) are created by stretching the heat map between 0 and 1 and multiplying this with the original image. We can clearly see that the trained network is very sensitive to local structures in the image. The presence of a secondary galaxy and/or disturbed features in parts of the primary galaxy (e.g. around the edges) all have a large impact in making a correct merger classification.  Figure ~\ref{heatmap_nmerger} shows examples of non-merging galaxies that were correctly identified by the CNN. The classifier is most sensitive to the features around the edges of the galaxy at the centre. Structures in the background  (e.g. artefacts from a bright source or an extended nearby object) do have some effect, but not large enough to change the classification result.

\end{appendix}

\begin{acknowledgements}
We would like to thank the anonymous referee for their insightful comments. We would like to thank the Center for Information Technology of the University of Groningen for their support and for providing access to the Peregrine high performance computing cluster.
GAMA is a joint European-Australasian project based around a spectroscopic campaign using the Anglo-Australian Telescope. The GAMA input catalogue is based on data taken from the Sloan Digital Sky Survey and the UKIRT Infrared Deep Sky Survey. Complementary imaging of the GAMA regions is being obtained by a number of independent survey programmes including GALEX MIS, VST KiDS, VISTA VIKING, WISE, Herschel-ATLAS, GMRT and ASKAP providing UV to radio coverage. GAMA is funded by the STFC (UK), the ARC (Australia), the AAO, and the participating institutions. Based on observations made with ESO Telescopes at the La Silla Paranal Observatory under programme ID 179.A-2004. Based on observations made with ESO Telescopes at the La Silla Paranal Observatory under programme IDs 177.A-3016, 177.A-3017, 177.A-3018 and 179.A-2004, and on data products produced by the KiDS consortium. The KiDS production team acknowledges support from: Deutsche Forschungsgemeinschaft, ERC, NOVA and NWO-M grants; Target; the University of Padova, and the University Federico II (Naples). 

\end{acknowledgements}

\end{document}